\documentclass[12pt]{amsart}
\usepackage{amssymb}
\usepackage{amsmath,amsthm,amsfonts}
\usepackage[applemac]{inputenc}
\usepackage{graphicx}
\usepackage{bbm}
\usepackage{pdfsync}
\usepackage{fancyhdr}
\usepackage{mathrsfs}
\usepackage{slashed}
\usepackage{color}
\usepackage{xcolor}
\usepackage[breaklinks,colorlinks,backref]{hyperref}

\hypersetup{
colorlinks, 
linktoc=all, 
linkcolor=blue, 
citecolor=hyptxt,
urlcolor=blue}
\hypersetup{
citebordercolor=,
filebordercolor=green,
linkbordercolor=blue
}
\definecolor{hyptxt}{rgb}{0.7, 0.4, 0.9}

\usepackage{bibtopic}

\usepackage[full]{textcomp} 
\usepackage{fbb} 
\usepackage[varqu,varl]{inconsolata}
\usepackage[libertine,bigdelims,vvarbb]{newtxmath}
\usepackage[cal=boondoxo]{mathalfa}
\usepackage[T1]{fontenc}
\useosf
\newcommand{\blue}[1]{\textcolor{blue}{#1}}

\newtheorem{prop}{Proposition}[section]

\newcommand{\beprop}{\begin{prop}}
\newcommand{\enprop}{\end{prop}}
\newcommand{\bprf}{\begin{proof}}
\newcommand{\eprf}{\end{proof}}

\newcommand{\ket}[1]{|\kern.3ex#1\kern.3ex\rangle}
\newcommand{\bra}[1]{\langle\kern.3ex #1 \kern.3ex|}
\newcommand{\scalar}[2]{\langle\kern.3ex #1 \kern.3ex|\kern.3ex#2\kern.3ex\rangle}

\def\R{\mathbb{R}}

\def\C{\mathbb{C}}

\def\ii{\mathrm{i}}
\def\ud{\mathrm{d}}

\def\ud{\mathrm{d}}
\definecolor{hervecolor}{rgb}{0.8,0,0.7}

\numberwithin{equation}{section}

\def\R{{\rm I\hspace{-.15em}R}}

\def\1{\mbox{I\hspace{-.15em}1}}

\def\setC{\mathbb{C}}

\def\z {\mathcal{ Z}}

\def\b{\begin{equation}}
\def\e{\end{equation}}
\def\tr{\mathrm{Tr}}
\begin{document}
\date{\today}
\title{Quantum Yang-Mills theory in \\ de Sitter ambient space formalism}
\author{M.V. Takook, J.P. Gazeau}

\address{\emph{ APC, UMR 7164}\\
\emph{Universit\'e de Paris} \\
\emph{75205 Paris, France}}

\email{ takook@apc.in2p3.fr, gazeau@apc.in2p3.fr}

{\abstract{
We present the quantum Yang-Mills theory in the four-dimensional de Sitter ambient space formalism. In accordance with the SU$(3)$ gauge symmetry the interaction Lagrangian is formulated in terms of interacting color charged fields in curved space-time. The gauge-invariant field equations are obtained in an independent coordinate description, and their corresponding color conserved currents are computed. Faddeev-Popov ghost fields are shown to be equivalent to their Minkowski counterparts. We obtain that the free ghost fields are massless minimally coupled scalar fields. The problems of the vacuum state, namely the breaking of de Sitter invariance, and the appearance of infrared divergence in its quantization procedure, are discussed. The existence of an axiomatic quantum Yang-Mills theory within the framework of the Krein space quantization is examined. The infrared divergence regularization of the interaction between the gauge vector fields and the ghost fields is studied in the one-loop approximation. Two different regularization methods are discussed: cut-off regularization and Krein space regularization. A mass term for the gauge vector fields is obtained, which may explain the mass gap and the color confinement problems at the quantum level in de Sitter background. The large curvature limit at the early universe or inflationary epoch is considered.}}

\maketitle

{\it Proposed PACS numbers}: 04.62.+v, 98.80.Cq, 12.10.Dm

\tableofcontents


\section{Introduction}
\label{intro}
Quantum field theory in de Sitter (dS) space-time has been evolving as an exceedingly vital subject in the past decade. Historically the dS space-time,
as a curved space-time manifold with maximum symmetry, was
introduced as a solution of Einstein's equations with a positive cosmological constant, ``maximum'' meaning that it has the same degree of symmetry as the flat Minkowski space solution \cite{bida}. The interest in the dS space has been tremendously increasing when it turned out that it could play a central role in the inflationary cosmological paradigm \cite{li}. Over the past three decades, a non-zero cosmological constant has been proposed to explain the luminosity of the farthest supernovae \cite{pe,plank}. Therefore the dS space-time could play a further important role in the modeling of the large-scale universe.

All of these developments make it more imperative than ever to seek a formulation of dS quantum field theory (QFT) with the same level of completeness and rigor as its Minkowskian counterpart. A unique Poincar\'e invariant vacuum state can be fixed
in Minkowski space by imposing the positive energy condition. In a curved space-time like de Sitter,
however, a global time-like Killing vector field does not exist, and therefore the positive energy condition cannot be imposed without any ambiguity.
Symmetry alone is not sufficient in the determination of a
unique vacuum state. Nevertheless, in dS space, symmetry considerations allow one to
identify the vacuum with a two-parameter ambiguity, say
$|\alpha,\beta\rangle$, corresponding to a family of distinct dS invariant vacuum states (see \cite{al} and references
therein). Only the one-parameter family $|\alpha,0\rangle$, is
invariant under the disconnected de Sitter group, {\it i.e.} O$(1, 4 )$ \cite{al}. By
imposing the condition that in the null curvature limit, the
Wightman's two-point function becomes the same as the Minkowskian
Wightman's two-point function, the other parameter $(\alpha)$, can be
fixed as well. This vacuum state, $|0,0\rangle \equiv |\Omega\rangle $, is called Bunch-Davies or Euclidean vacuum state. However, this vacuum state is not suitable when one deals with the massless minimally coupled (\textit{mmc}) scalar fields, which causes problems for quantization of the Yang-Mills theory and the linear gravity.

In \cite{brgamo,brmo} Bros et al. have presented a QFT for a massive scalar field in
dS space-time that closely mimics the QFT in Minkowski space-time. They
have introduced a new version of the so-called Fourier-Bros transformation
on the dS hyperboloid, which allows us to completely
characterize the Hilbert space of the ``one-particle'' states in terms of
corresponding irreducible unitary representations of the dS
group. In this construction, the analyticity properties of the two-point functions play a significant role in the interaction field quantization. Then the Bros construction was generalized to various quantum free fields with non-zero spin in dS space in a series of papers (see \cite{ta97} and references therein).

On the other hand, quantum Yang-Mills theory can be considered one of the most mathematically challenging topics in QFT \cite{jawi}.
An axiomatic quantum Yang-Mills theory, with a mass gap and color confinement, does not yet exist\footnote{It is one of the seven Millennium Prize Problems in mathematics that were stated by the Clay Mathematics Institute in May $24, 2000$ \cite{jawi}.}. Recently, a dS Yang-Mills theory was studied in an intrinsic coordinate system \cite{cklp}. In the present paper, the Yang-Mills theory is considered within the $4$-dimensional dS ambient space formalism. We describe the interaction between the color-charged spinor field (``quark'') and the massless color-charge vector fields (``gluons'') in a coordinate-independent way. We obtain the gauge-invariant Lagrangian, the field equations, and the gauge fixing terms in ambient space formalism. We then show that the free ghost fields are the \textit{mmc} scalar fields in dS space-time.

The famous infrared (IR) divergence of the \textit{mmc} scalar field and the different ways of dealing with it were addressed by many authors over the last $30$ years. We will briefly recall the crucial results. It was commonly thought that this IR divergence is just a gauge artifact \cite{anilto}. The appearance of this IR divergence in the Faddeev-Popov ghost fields of dS perturbative gravity in the covariant gauge has also been noticed (for instance, see \cite{gihili}). The effect of an IR cut-off in the momentum space, which regularizes this divergence, breaks the dS invariance. About this problem, we will here insist on the necessity of using the Krein space quantization to preserve the dS invariance, as was shown in the covariant quantization of the \textit{mmc} scalar fields in dS space \cite{dere98,gareta,gahure}.

The Krein approach was implemented for the massless fields with spin $1, 3/2$, and $2$ with the construction of relevant indecomposable representations \cite{gagarota,paenta,pegazeau}. The content of the present paper also rests upon the results of \cite{brgamota,gata,taazba,pegazeau2,gagata} which were based on the axiomatic quantum field theory. The Krein space quantization may be used to construct an axiomatic dS quantum Yang-Mills theory on the Krein-Fock vacuum state.

The interaction of the ghost field and the gauge vector field at the QFT level is discussed in the one-loop approximation. We show that, due to the regularization of the IR divergence, a mass term appears for gluons fields, which may explain the mass gap and color confinement problems. The regularization of the IR divergence is considered along two different paths: A) a cut-off regularization method \cite{al}, and B) the Krein space regularization method \cite{ta02,fotaza,ta05,khnarota,reta}. The first one breaks the dS invariance while the second one preserves it.

At this point, it is necessary to recall that the concept of (proper) mass is defined within the context of the  Poincar\'e group symmetry, and has to be carefully revisited within the context of the dS group symmetry. In the present paper, a field is called ``massive'' when it propagates inside the light-cone and corresponds to a massive Poincar\'e field representation in the null curvature limit. We call a field ``massless'' if it propagates on the dS light-cone and corresponds to a massless Poincar\'e field representation at $H=0$. The concept of light-cone propagation can be precisely defined in dS ambient space formalism \cite{brmo}.

The organization of this paper is as follows. In Section \ref{notations} we fix the notations, and we make explicit the field equations for Dirac and massless vector fields, the Lagrangian density, the conserved current, and finally, the massless vector two-point function. In Section \ref{NAGT} we obtain the non-abelian gauge-invariant Lagrangian. Then the invariant field equations and their corresponding color conserved currents are computed. By obtaining the Faddeev-Popov ghost fields, it is shown that these fields are \textit{mmc} scalar fields. In Section \ref{mmcfield} the problems of the vacuum state, the breaking of the Sitter invariance, and the appearance of infrared divergence in the quantization procedure are reviewed. Section \ref{AxQT} is devoted to the quantum Yang-Mills theory, and two different approaches are presented. The mass gap and the color confinement problems of the Yang-Mills theory are discussed in Section \ref{gapcolor} through the consideration of the one-loop effective action corresponding to the gluon-ghost interaction. Finally, concluding remarks and a brief outlook are given in Section \ref{conclu}.


\section{Notations} \label{notations}

The dS space-time can be identified with the $4$-dimensional hyperboloid embedded in the $5$-dimensional Minkowski space-time as:
\b \label{dSs} X_H=\left\{ x^\alpha\equiv x \in \R^5| \; \; x \cdot x=\eta_{\alpha\beta} x^\alpha
x^\beta =-H^{-2}\right\}\,,\;\; \alpha,\beta=0,1,2,3,4\,, \e
with $\eta_{\alpha\beta}=$diag$(1,-1,-1,-1,-1)$ and $H$ is like Hubble's constant parameter. The dS metric element is:
\b \label{dsmet} \ud s^2=\left.\eta_{\alpha\beta}\ud x^{\alpha}\ud x^{\beta}\right|_{x\cdot x=-H^{-2}}=
g_{\mu\nu}^{dS}\ud X^{\mu}\ud X^{\nu}\,,\;\; \mu=0,1,2,3\,,\e
where the $X^\mu$'s form a set of $4$-space-time intrinsic coordinates on the dS hyperboloid, and the $x^{\alpha}$'s are the ambient space coordinates.
Let us introduce the global coordinates system in terms of the intrinsic coordinates system $(X^\mu)=(t,\chi,\theta,\varphi)$, $t\in\R$, $0\leq\chi,\theta\leq\pi$, $0\leq\varphi<2\pi$ as:
\b \label{gcs} \left\{\begin{array}{clcr} x^0&=H^{-1}\sinh Ht \\
x^1&=H^{-1}\cosh Ht\sin\chi \cos\theta\\
x^2&=H^{-1}\cosh Ht\sin\chi \sin\theta\cos\phi \\
x^3&=H^{-1}\cosh Ht\sin\chi\sin\theta\sin\phi\\
x^4&=H^{-1}\cosh Ht\cos\chi \, ,
\end{array} \right.\e
which are appropriate to take the zero curvature limit. The analyticity properties of QFT in dS space are described with the introduction of the complexified de Sitter spacetime $X_H^{(c)}$:
$$ X_H^{(c)}=\left\{ z=x+\ii y\in \C^5;\;\;\eta_{\alpha \beta}z^\alpha z^\beta=(z^0)^2-\vec z.\vec z-(z^4)^2=-H^{-2}\right\}$$
\b =\left\{ (x,y)\in \R^5\times \R^5;\;\; x^2-y^2=-H^{-2},\; x.y=0\right\}\,.\e
The forward and backward tubes in $ \C^5$ are defined as $T^\pm= \R^5+\ii V^\pm$. The domain $V^+$(resp. $V^-)$
stems from the causal structure on $X_H$:
\b V^\pm=\left\{ x\in \R^5;\;\; x^0 \gtrless \sqrt {\parallel \vec x\parallel^2+(x^4)^2}\right \}\,.\e
Their respective intersections with $X_H^{(c)}$ are denoted by:
\b \mathcal{T}^\pm=T^\pm\cap X_H^{(c)}\,,\e
and will be called forward and backward tubes of the complex dS space, respectively. Finally
the ``tuboid'' above $X_H^{(c)}\times X_H^{(c)}$ is defined by \cite{brmo}:
\b \mathcal{T}_{12}=\{ (z,z');\;\; z\in \mathcal{T}^+,z' \in \mathcal{T}^- \}\,. \e

In the ambient space formalism, the tangential derivative on the dS hyperboloid reads as
$
\partial_\beta^\top =\theta_{\alpha \beta}\partial^{\alpha}=
\partial_\beta + H^2 x_\beta\, x\cdot\partial\,,
$
where $ \theta_{\alpha \beta}=\eta_{\alpha \beta}+
H^2x_{\alpha}x_{\beta}$
is the transverse projector.
The transverse-covariant derivative acting on a tensor field of rank-$l$ is defined by:
\begin{equation}\label{dscdrt}
\nabla^\top_\beta T_{\alpha_1\cdots\alpha_l}\equiv \partial^\top_\beta
T_{\alpha_1\cdots\alpha_l}-H^2\sum_{n=1}^{l}x_{\alpha_n}T_{\alpha_1\cdot
\alpha_{n-1}\beta\alpha_{n+1}\cdots\alpha_l}\,.
\end{equation}
The second-order Casimir operator $Q_0$ of the dS group SO$_0(1,4)$ in its representation acting on scalar fields is written as: \begin{equation}
\label{casQ0}
Q_0=-H^{-2}\partial^\top\cdot\partial^\top=-H^{-2}g^{dS}_{\mu\nu}\nabla^\mu\nabla^\nu=-H^{-2} \Box_H \,,
\end{equation}
where $ \Box_H$ is the Laplace-Beltrami operator on the dS space-time.

By expressing the Casimir eigenvalue equation in terms of the
infinitesimal generators of the group representation and proceeding with a dS-Dirac field factorisation of the quadratic terms, one obtains the first order equation \cite{brgamota,dir}:
\b \label{dsdfe} \left(-\ii H\slashed{x}\gamma\cdot \partial^\top
+\ii 2H+H\nu\right)\psi(x)=0,\;\; \nu \in \R,\;\;\slashed{x}=x\cdot \gamma\,,\e
where $\nu$ is the parameter of the principal series representation, as was introduced in \cite{brgamota}. This equation can be derived from the least action principle applied to the following spinor field action:
\b \label{dsdlag} S[\psi,\bar{\psi}]=\int \ud\sigma(x) \mathcal{ L}(\psi,\bar{\psi})=\int \ud\sigma(x) \bar
\psi\gamma^4\left(-\ii H\slashed{x}\gamma\cdot{\partial}^\top+\ii 2H+H\nu\right)\psi(x)\,,\e
where $\bar{\psi}=\psi^\dag \gamma^0\gamma^4$ and $\ud\sigma(x)$ is the $4$-dimensional dS-invariant volume element. The latter is  expressed in the ambient space formalism as \cite{brmo}:
\b \label{mainds} d\sigma(x)=\left.\frac{dx^{0}\wedge dx^{1}\wedge dx^{2}\wedge dx^{3} \wedge dx^{4}}{d(x \cdot x+H^{-2})}\right|_{X_H}=d\sigma (x')\, .\e
Here we need five
$\gamma^\alpha$ matrices instead of the usual four ones in the Minkowski
space-time. They generate the Clifford algebra associated with the five dimensional Minkowskian metric as:
$$ \{ \gamma^\alpha ,\gamma^\beta \}=\gamma^\alpha
\gamma^\beta+\gamma^\beta \gamma^\alpha =
2 \eta^{\alpha \beta}\;\;\;,\;\;\;
\gamma^{\alpha\dag}=\gamma^0 \gamma^\alpha \gamma^0\,.$$
By using the Noether's theorem and the invariance of the field equation (\ref{dsdfe}) under the global $U(1)$ transformation ($\psi'=e^{-\ii \Lambda}\psi,\, \Lambda=$ constant), the conserved current can be obtained \cite{rota05}:
$$ J^\alpha(x )= \frac{\delta \mathcal{ L}(\psi,\bar{\psi})}{\delta
\partial^{\top}_\alpha \psi} \frac{\delta \psi}{\delta \Lambda
}=-H\bar \psi(x)\gamma^4\slashed{x} \gamma^\alpha \psi(x)\,.$$
It is easily verified that this current satisfies the
conditions \cite{brgamota,rota05}:
\b \label{ccabelin} \partial^\top\cdot J(x)=0\,,\;\;\;\; x\cdot J(x)=0\,.\e

For the massless vector field (spin $=1$), one also starts from the Casimir operator and its eigenvalue, uses the infinitesimal
generators and some algebraic relations, to finally obtain the following field equations \cite{gata,gagarota}:
\b \label{maveds}\left(Q_0-2\right)K_\alpha(x)+2x_\alpha \partial^\top\cdot K(x)+H^{-2} \partial^\top_\alpha \partial^\top \cdot K=0\,.\e
This equation is invariant under the  gauge transformation:
\b \label{vgintr} K_\alpha \longrightarrow K'_\alpha=K_\alpha+H^{-2} \partial^\top_\alpha \Lambda(x) \,.\e
Here $\Lambda(x)$ is an arbitrary differentiable scalar field. The condition of transversality $ x\cdot K(x)=0 $ restricts the five-vector field to live on the dS hyperboloid and guarantees that it should be viewed as a vector-valued homogeneous function of the $\R^5$-variables $x^{\alpha}$ with some arbitrarily chosen degree $\lambda$ \cite{dir}:
\b x^{\alpha}\frac{\partial }{\partial
x^{\alpha}}K_{\beta}(x)=x\cdot \partial \,K_\beta (x)=\lambda
K_{\beta}(x)\,. \e

The massless vector field equation can be derived from the following action integral:
\b \label{guvela2} S[K]=-\frac{1}{4} \int \ud\sigma(x)\left( \nabla^\top_\alpha K_\beta-\nabla^\top_\beta K_\alpha\right)\left(\nabla^{\top\alpha} K^{\beta}-\nabla^{\top\beta} K^{\alpha}\right).\e
Like for its flat space counterpart, the gauge fixing is accomplished by adding to (\ref{maveds}) a gauge
fixing term with parameter $\zeta$ \cite{gagarota}:
\b \label{gfmaveds}\left(Q_0-2\right)K_\alpha(x)+2x_\alpha \partial^\top\cdot K(x)+\zeta H^{-2} \partial^\top_\alpha \partial^\top \cdot K=0\,.\e
 With  the simple choice $\zeta=\frac{3}{2}$, the vector two-point function reads \cite{gagarota}:
\b \label{vtfsg} \mathcal{ W}_{\alpha \alpha'}(x,x')=\langle
\Omega\mid K_{\alpha}(x)K_{\alpha'}(x')\mid \Omega \rangle= D_{\alpha
\alpha'}(x,x',\partial_{x})\mathcal{ W}_{mcc}(x,x')\,,\e
where $D_{\alpha \alpha'}(x,x',\partial_{x})$ is the following bi-tensor differential operator,
\b \label{do} D_{\alpha \alpha'}(x,x',\partial_{x})= \theta_{\alpha }\cdot\theta'_{\alpha' }+H^{-2}
\partial^\top_\alpha \left[ \partial^\top \cdot \theta'_{\alpha' }+H^2x \cdot\theta'_{\alpha'
}\right].\e
The symbol $\mathcal{ W}_{mcc}(x,x')$ stands for the Wightman two-point function of the massless conformally coupled scalar field \cite{brmo,chta68,tag}:
\b \label{stpci2}
\mathcal{ W}_{\mathrm{mcc}}(x,x')=\frac{-H^2}{8\pi^2}\left[
P\frac{1}{1-\mathcal{ Z}(x,x')}
-\ii\pi\epsilon(x^0,x'^0)\delta(1-\mathcal{ Z}(x,x'))\right]\,.\e
Here the symbol $P$ means the principal part and $$ \epsilon (x^0-x'^0)=\left\{\begin{array}{clcr} 1&x^0>x'^0
\\
0&x^0=x'^0\\ -1&x^0<x'^0\\ \end{array} \right. .$$
The function $\mathcal{ Z}$ is given by:
\b \label{gedi} \mathcal{ Z}(x,x')=-H^2 x\cdot x'=1+\frac{H^2}{2} (x-x')^2 \equiv \cosh H
\sigma (x,x')\, ,\e
and $\sigma (x-x')$ is the geodesic distance between two points $x$ and $x'$ on the dS hyperboloid.


\section{Lagrangian density}
\label{NAGT}

In this section, a direct generalization of the abelian gauge theory expressed in terms of the dS ambient space \cite{rota05} to the non-abelian gauge theory is presented. In this formalism, the gauge theory construction is utterly similar to its Minkowski counterpart \cite{wei2}. The dS-Dirac spinor field equation (\ref{dsdfe}) is not invariant under the following non-abelian gauge transformation:
\b \psi'(x)=e^{-\ii\Lambda^a(x)t^a}\psi(x)\equiv U\left(\Lambda(x)\right)\psi(x)\,. \e
The $\Lambda^a$'s are local group parameters and the $t^a$'s are generators of the $\mathrm{SU}(3)$ group. They satisfy the following commutation relation:
\b [t_a,t_b]=\ii C_{ab}^{\;\; \;c}t_c, \;\; a,b,c=1,2,\cdots, 8\,,\e
where $C_{ab}^{\;\; \;c}$'s are the real structure constants of $\mathfrak{su}(3)$. 

Like the flat Minkowskian case, local gauge symmetry with its space-time-dependent transformation can generate the gauge interaction in the dS universe. For obtaining a gauge-invariant Lagrangian for spinor field, it is necessary to replace the transverse-covariant derivative $ \nabla^\top_\beta$ with the gauge-transverse-covariant derivative $D_\beta^{K}$ which is defined by:
\b D_\beta^K \equiv\nabla^\top_\beta -\ii g K_{\beta}^{a}t_a\,\, , \;\; x\cdot K^{a}=0\, ,\e
where $g$ is a real coupling constant. The gauge fields $K_{\beta}^a(x)$ are associated with the group generators $t^a$.
In order to have a simple gauge transformation for the gauge-covariant derivative of the spinor field as:
\b D_\beta^K\psi(x)
\longrightarrow \left[D_\beta^{K}
\psi(x)\right]'=e^{-\ii\Lambda^a(x)t^a}D_\beta^{K}\psi(x)\equiv D_\beta^{K'}
\psi'(x)\, , \quad D_\beta^{K'}=UD_\beta^{K} U^{-1}\, ,\e
the connection fields or vector gauge potentials $K_\beta^{a}$ must be transformed in the following form:
\b \label{nagtofk} {K^{\prime}}_\beta^{a}t_a=U (\Lambda) K_\beta^{b}t_b U^{-1}(\Lambda)+\frac{\ii}{g} U\left(\partial^\top_\beta U^{-1}\right) \,.\e
Then the dS-Dirac gauge invariant equation reads as:
\b \left(-\ii H\slashed{x}\gamma.D^K +2H\ii+H\nu\right)\psi(x)= \left(-\ii H\slashed{x}\gamma\cdot{\partial}^\top-Hg\slashed{x}\gamma\cdot K^a t_a +2H\ii+H\nu\right)\psi(x)=0\,.\e

By using the curvature $\mathcal{C}$ of the Lie algebra of the gauge group $\mathrm{SU}(3)$, one obtains the field equations for the vector fields $K_\beta^{a}$. The gauge group curvature is:
$$\mathcal{C}\left(D^K_\alpha,D^K_\beta\right)=\frac{\ii}{g}\left[D^K_\alpha,D^K_\beta\right] = F_{\alpha\beta}^{\;\;\;\;a}t_a \equiv\, {\bf F}_{\alpha\beta}\,,\;\;\; {\bf F}_{\alpha\beta}'=U{\bf F}_{\alpha\beta}U^{-1}\, ,$$
where
\b F_{\alpha\beta}^{\;\;\;\;a}=\nabla^\top_\alpha K_\beta^{\;\;a}-\nabla^\top_\beta K_\alpha^{\;\;a}+ gC_{bc}^{\;\;\;\;a}K_\alpha^{\;\;b}K_\beta^{\;\;c}\,,\e
together with the transversality properties
$x^\alpha F_{\alpha\beta}^{\;\;\;\;a}=0= x^\beta F_{\alpha\beta}^{\;\;\;\;a}$.
The $\mathrm{SU}(3)$ gauge invariant action or Lagrangian in the dS background for the gauge field $K_\alpha^{\;\;a}$ is:
$$ S[K^a]=-\frac{1}{2}\int \ud \sigma(x) \; \tr\left({\bf F}_{\alpha\beta}{\bf F}^{\alpha\beta} \right)=-\frac{1}{2}\int \ud\sigma(x) F_{\alpha\beta}^{\;\;\;\;a}F^{\alpha\beta b}\, \tr\left(t_at_b\right)\, , $$
where summing over the repeated indices is used. The normalization of the structure constants is usually
fixed by requiring that, in the fundamental representation, the corresponding
matrices of the generators $t_a$ are normalised such as \cite{ilio13}
$ \tr\left(t_at_b\right) =\frac{1}{2} \delta_{ab}$. Then the action becomes
$$ S[K^a]=-\frac{1}{4}\int \ud \sigma(x)\left[ \left( \nabla^\top_\alpha K_\beta^{\;\;a}-\nabla^\top_\beta K_\alpha^{\;\;a}\right)\left(\nabla^{\top\alpha} K^{\beta b}-\nabla^{\top\beta} K^{\alpha b}\right)\delta_{ab}+O({K}^3)\right]\,,$$
where non-linear terms in the field equation appear in $O(K^3)$ and describe the interaction between the gauge potentials $K_\beta^{a}$. One can see that the gauge potential or connection $K_{\beta}^{a}$ in the linear approximation is a massless vector field with action integral (\ref{guvela2}).

The gauge invariant Lagrangian density for the vector and spinor fields in the simplest form can be written as:
\b \mathcal{ L}(K,\psi)=-\frac{1}{4}F_{\alpha\beta}^{\;\;\;\;a}F^{\alpha\beta a}+H\bar
\psi\gamma^4\left(-\ii\slashed{x}\gamma\cdot D^{K}+2\ii +\nu \right)\psi\, .\e
From the Euler-Lagrange equations one obtains \cite{ratapol}:
\b \nabla^\top_\alpha F^{\alpha\beta}_{\;\;\;\;a}=gF^{\beta\gamma b} C_{ba}^{\;\;\;c} K_{c\gamma}
+g H\bar \psi(x)\gamma^4\slashed{x} \gamma^\beta t_a\psi(x)\, ,\e
\b \left(-\ii H\slashed{x}\gamma\cdot{\partial}^\top +2H\ii +\nu H\right)\psi(x)=gH\slashed{x}\gamma \cdot K^a(x)
t_a\psi(x)\, .\e
In this case the conserved current reads as:
\b \mathcal{ J}_\alpha^{ a}= J_\alpha^{a (K)} + J_\alpha^{a (\psi)}=-gF_{\alpha \beta b}C^{ba}_{\;\;\; c}K^{c\beta} -g H\bar \psi_i(x)\gamma^4\slashed{x} \gamma_\alpha (t^a)_{ij}\psi_j(x)\, ,\e
where $i,j=1,2,3$ are the colour indexes of spinor fields. By using the equation (\ref{ccabelin}), one verifies that this current satisfies the following conditions:
\b x\cdot \mathcal{ J}^a=0\,, \; \nabla^\top\cdot \mathcal{ J}^a=0=\partial^\top\cdot \mathcal{ J}^a\,.\e
By comparing this current with the abelian case (\ref{ccabelin}), we see that the term $J_\alpha^{a (K)}$ is due to the color vector fields in dS space-time.

The two conditions of transversality, $x\cdot K=0$, and divergencelessness, $\nabla^\top\cdot K=0$, are combined in one mathematical relation, $\partial^\top\cdot K=0$. It is called the \emph{generalized Lorenz gauge}, which will be considered in this paper. For the infinitesimal gauge transformation, the equation (\ref{nagtofk}) becomes:
\b (K^\Lambda)_\alpha^a=K_\alpha^a+ \frac{1}{g} \partial_\alpha^\top \Lambda^a+C_{cb}^{\;\;a}K^b_\alpha\Lambda^c\, ,\e
and the generalized Lorenz gauge is rewritten as:
\b \label{gffe} \chi^a=0=\partial^\top \cdot (K^\Lambda)^a=\partial^\top \cdot K^a-\frac{H^2}{g}Q_0\Lambda^a+C_{cb}^{\;\;a}\partial^\top \cdot K^b \Lambda^c\;.\e
This equation fixes the gauge parameter through the equation:
\b \label{gfghost} \left(\delta^{ac}H^2Q_0-gC_{cb}^{\;\;a}\partial^\top \cdot K^b\right) \Lambda^c=0\,. \e
Like its Minkowskian counterpart, the Faddeev-Popov ghost procedure can be implemented by using the expression:
$$ \det M_{ab}=\det \left|\frac{\delta \chi^{ a}(y)}{\delta \Lambda^b(x)} \right| =\det \left(\delta^{ab}H^2Q_0-gC_{cb}^{\;\;a}\partial^\top \cdot K^c\right)\,. $$
Let us introduce the ghost field $\Phi$ to absorb this factor in the Lagrangian density. The Lagrangian density of the ghost field for the generalized Lorenz gauge is obtained in the same way as its Minkowskian counterpart. By using the equation (\ref{gffe}) and the path integral formulation, the Lagrangian density of the ghost field in the generalized Lorenz gauge is found to be:
\b \mathcal{ L}(\Phi, \Phi^\dag)={\Phi^a} ^\dag\left[ \delta^{ac}H^2Q_0-gC_{cb}^{\;\;a}\partial^\top \cdot K^b \right] \Phi^c(x)\,.\e

It is important to note that the field equation for the free ghost fields is the same as for the \textit{mmc} scalar field. Within the dS context, the scalar free ghost fields transform analogously to the \textit{mmc} scalar field \cite{gareta} under the action of an indecomposable representation of the dS group. Concerning the gauge group indices $a$ for $\Phi^a$, they transform under the adjoint representation analogously to the gluon fields. The many ``particle'' states are constructed from the Fermi-Dirac statistics resembling its Minkowskian counterpart.

The gauge fixing classical Lagrangian density for the interaction between colour charged particles reads as:
 \begin{equation}
\label{spvegh} 
 \begin{split}
\mathcal{ L}_c(\psi,\bar \psi, K^a, \Phi,\Phi^\dag)&= H\bar
\psi\gamma^4\left(-\ii\slashed{x}\gamma\cdot D^{K}+2\ii +\nu \right)\psi \\
& -\frac{1}{4}F_{\alpha\beta}^{\;\;\;\;a}F^{\alpha\beta a}-\frac{\zeta}{2} \partial\cdot K^{\;a}\partial\cdot K^{\;a} +{\Phi^a} ^\dag\left[ \delta^{ab}H^2Q_0- gC_{cb}^{\;\;a}\partial^\top \cdot K^c \right] \Phi^b\,,
 \end{split}
\end{equation}
where $\zeta$ is a gauge fixing parameter. The spinor fields correspond to the quarks with three colors, and the vector fields correspond to the gluons with eight combinations of the three colors. Eight colors ghost fields are also presented. In the quantization of the ghost field appears an IR divergence in the propagator, and its regularization, thanks to a cut-off (see \cite{alfol}),  breaks the dS invariance. The \textit{mmc} scalar field is invariant under the global transformation ($\phi$ + const.). On the quantum level this feature  is equivalent to a gauge theory \cite{gareta} and also to the two-dimensional QFT \cite{ford}.


\section{Quantization problems}
\label{mmcfield}

For the quantization of the Yang-mills theory, {\it i.e.} quantization of the Lagrangian density (\ref{spvegh}), first, the quantization of the free fields are considered, and then, by using the perturbation theory, the quantization of the interaction part is presented. For the free field quantization, the propagator has to be calculated. For the interaction field, the renormalized effective action is a significant quantity that must be obtained. There are three types of free fields in the Yang-Mills theory: spinor fields, massless vector fields, and \textit{mmc} scalar fields. The quantization of the spinor field can be performed without any problem. The quantization for the vector field is discussed in the next section. In this section, we describe the quantization of the \textit{mmc} scalar fields.

In the quantization procedure for the \textit{mmc} scalar field, three problems emerge that are related to each other: (1) appearance of IR divergence, (2) breakdown of dS background symmetry, and (3) absence of de Sitter-invariant Fock vacuum state \cite{al,alfol}. These difficulties arise from the zero-mode problem associated with a global transformation of the field equation ($\phi'=\phi$ + const.). Similar problems exist in quantum gauge theory.  An indefinite metric quantization can be used to solve them \cite{stroch}. In the case of the \textit{mmc} scalar field the so-called Krein space quantization, which was developed in \cite{dere98,gareta}, represents a consistent alternative to overcome these obstacles: removing the infrared divergence and preserving the dS invariance. Concerning the third problem, the alternative is to introduce the Krein-Fock vacuum state as a workable remedy.

As proved by Allen \cite{al}, the covariant canonical quantization
procedure with positive norm states fails in the \textit{mmc} case. The Hilbert space generated by the positive modes, including the zero mode ($\phi_{0}$), is not de Sitter invariant \cite{al,gareta}:
$$\mathcal{ H}=\left\{\sum_{k \geq 0}\alpha_k\phi_k;\;
\sum_{k \geq 0}|\alpha_k|^2<\infty\right\}\,.$$
It means that the positive modes solutions are not closed under the action of the de Sitter group generators. Nevertheless, one can obtain a fully covariant quantum field operator by adopting
a new construction, which consists in adding all the conjugate modes to the original ones.
Consequently, one now deals with an orthogonal sum of a positive and
a negative inner product space, which is closed under an indecomposable
representation of the dS group \cite{gareta}.

Since the spinor field and vector field are studied in ambient space formalism, and for the unification of the formalism, the \textit{mmc} scalar field is presented in ambient space formalism, which permits us to better understand the breaking of dS invariance. In the ambient space formalism one easily constructs the \textit{mmc} scalar field $\Phi_{\mathrm{mmc}}$ in terms of the massless conformally coupled scalar field (\textit{mcc}) $\Phi_{\mathrm{mcc}}$, and an arbitrary constant five-vector $A=\left(A^\alpha\right)$ \cite{ta97,gatahi}:
\begin{equation}
\label{mmcmcc}
\Phi_{\mathrm{mmc}}(x)\equiv \left[A\cdot\partial^\top + 2H^2 A\cdot x\right]\Phi_{\mathrm{mcc}}(x)\, ,
\end{equation}
It is reminiscent of the arbitrariness in the choices of the vacuum state. The \textit{mcc} scalar field two-point function is constructed on the Bunch-Davies vacuum state (\ref{stpci2}).

Using the equation (\ref{mmcmcc}), the analytic (in its domain) two-point function for the massless minimally coupled scalar field can be written as:
\b \label{mth} W_{\mathrm{mmc}}(z,z';A) =\left[A\cdot\partial^\top + 2H^2 A\cdot z\right]\left[A\cdot\partial'^\top + 2H^2 A\cdot z'\right]W_{\mathrm{mcc}}(z,z') \, ,\e
where $z =x+\ii y$, $z'=x'+\ii y'$ and $W_{mcc}$ is the analytic two-point function of the \textit{mcc} scalar field \cite{brmo}:
$$ W_{mcc}(z,z')=\frac{H^2}{8\pi^2}\frac{-1}{1-\mathcal{ Z}(z,z')}\, . $$
The Wightman two-point function $\mathcal{ W}_{\mathrm{mcc}}(x,x')$, Equation (\ref{stpci2}), is the boundary value of the analytic function $W_{\mathrm{mcc}}(z,z')$ (in the distributional sense, according to the theorem $A.2$ in \cite{brmo}).

After some calculations, one obtains from \eqref{mth} \cite{gatahi}:
\begin{equation}
\label{Wmmc}
\begin{split}
&W_{\mathrm{mmc}}(z,z';A)=\frac{-H^2}{8\pi^2} \times \\
&\frac{(\mathcal{ Z}-3)\left[(H^2A\cdot z)^2 +(H^2A\cdot z')^2+H^4A\cdot z\, A\cdot z' \,\mathcal{ Z}\right]
+6H^4A\cdot z A\cdot z'-(1-\mathcal{ Z})H^2A\cdot A }{\left[1-\mathcal{ Z}(z,z')\right]^3}\, .
\end{split}
\end{equation}
This function is not dS invariant concerning the variables $z$ and $z^{\prime}$. Instead, we have to consider the $A$-labeled family of two-point functions (or ``vacuum state'') for which the following dS invariance holds:
\begin{equation}
\label{Ainvar}
W_{\mathrm{mmc}}(R z, R z';R A)= W_{\mathrm{mmc}}(z,z';A)\quad \mbox{for all} \ R \in \mathrm{SO}_0(1,4)\, .
\end{equation}
This means that it is built from a one-particle state which does not transform under the unitary irreducible representation of the dS group. $A$ is a constant five-vector. It is a sort of polarization vector \cite{gaha88} which can be chosen as one of the $5$ vectors forming an orthonormal basis of $\R^5$, the latter carrying the fundamental five-dimensional representation of the dS group. Hence, the explicit form of the two-point function $W_{mmc}$ depends on the chosen $A$. Its construction involves the tensor product of two representations of the dS group: $(1)$ the complementary scalar representation related to the massless conformally coupled scalar field \cite{ta97}, and $(2)$ the fundamental five-dimensional representation \cite{gaha88}.

As a first simple example of a choice of orthonormal basis in $\R^5$ (for the dS metric) , one considers the set $\{A^{(l)}\, , \, l=0,1,2,3,4\}$ obeying \cite{gaha88}:
\b \label{zpolar} \sum_{l=0}^4 \sum_{l'=0}^4 A^{(l)}_\alpha A^{(l')}_\beta=\eta_{\alpha\beta}\,,\;\;\; A^{(l)}\cdot A^{(l')}=\eta ^{ll'}\, .\e
With this choice and by summing the $5$ corresponding two-point functions one obtains the constant trivial solution (for $z\neq z'$) \cite{gatahi}:
\b \label{consolu} W_{mmc}(z,z')= \left[\partial^\top\cdot\partial'^\top +2H^2z\cdot\partial'^\top + 2H^2 z' \cdot \partial^\top+ 4 H^4z\cdot z'\right] W_{mcc}(z,z')=\frac{-H^2}{8\pi^2}\, .\e
It is precisely the constant solution obtained by Allen \cite{al}. In this case, we have restored the trivial SO$(1,4)$ invariance, but then we meet the problem of the non-existence of a Fock vacuum state \cite{al}.

A second example is provided by the elementary choice
\b \label{so4inv} A\equiv (1,0,0,0,0)\, .\e
We then obtain the O$(4)$ invariant two-point function:
\begin{equation}
\label{ }
W_{\mathrm{mmc}}(z,z') =\frac{-H^2}{8\pi^2}
\frac{(\mathcal{ Z}-3)\left[(H^2z^0)^2 +(H^2 z'^0)^2+ H^4z^0\, z'^0 \,\mathcal{ Z}\right]+6 H^4 z^0 z'^0-H^2(1-\mathcal{ Z}) }{\left[1-\mathcal{ Z}(z,z')\right]^3}\, .
\end{equation}
This two-point function is free of logarithmic divergence but it breaks the dS invariance, as expected. This result was previously discussed by Allen-Folacci in \cite{al,alfol}. As a matter of fact, the Allen-Folacci expression has the following logarithmic divergence:
\b \label{alfol} \mathcal{ W}^{\mathrm{AF}}_{mmc}(x,x') \propto \left[\frac{1}{1-\mathcal{ Z}}-\ln (1-\mathcal{ Z})+ \cdots\right]\,.\e
Going back to the general expression \eqref{Wmmc} for $W_{\mathrm{mmc}}$, its dominant term at large values of $\mathcal{ Z} \sim (z-z^{\prime})^2\sim -z\cdot z^{\prime}\longrightarrow \infty$, reads as
$$-\frac{H^4}{8\pi^2}\left[\frac{A\cdot z\, A\cdot z' }{ z\cdot z^{\prime}}+ \frac{(A\cdot z)^2 + (A\cdot z^{\prime})^2}{ (z\cdot z^{\prime})^2}\right] := f(z,z^{\prime},A)\, . $$
One observes that at large separated points, we get a $A$ dependent value, which generally depends on the respective directions of $z$ or $z'$ with respect to $A$. As an illuminating example, we fix $z^{\prime}=(0,0,0,0,H^{-1} )$ and we choose $z=H^{-1}(\sinh Ht ,0,0,0, \cosh Ht )$, with $ t \longrightarrow \infty$, then we obtain: $\lim_{z\to \infty} f=-\frac{H^4}{4\pi^2}(A_4)^2$. Importantly, $A$ might be related to the cut-off introduced by Allen in \cite{al}. Also, might this behaviour explain the appearance of a mass, analogous to the Higgs mechanism? 

In contrast, we assert that within the framework of Krein space quantization, as is described below in Subsection \ref{kreinapp}, de Sitter invariance is preserved, as exemplified with Equation (\ref{tfsmin}), and we obtain a constant term for large separated points.


\section{Quantum Yang-Mills theory}
\label{AxQT}

In QFT, the two-point function and the time-ordered product propagator are the significant quantities that must be calculated. The first one is used for the construction of an axiomatic quantization \cite{stwi} and the second one for the calculation of the effective action or S-matrix element in the perturbation theory \cite{bailse}. It is well known that for a quantum gauge theory, an indefinite metric quantization must be used \cite{stroch}. The covariant quantization of the massless vector field and \textit{mmc} scalar field require an indecomposable representation of the de Sitter
group. The indecomposable representations are not unique and depend on gauge fixing parameters. The existence of an axiomatic quantization of the quantum Yang-Mills theory is considered in two different approaches.

In the first approach, the positive frequency solution of the field equation is used, and it is called the standard approach. In this approach, the field operator is defined with the help of coordinate-independent dS plane waves (modes). The construction is based on the analyticity requirements in the complexified pseudo-Riemannian manifold. The field equation's positive and negative frequency solutions are used in the second approach, based on Krein space quantization. It is called the Krein approach in this paper.

\subsection{Standard approach}

The axiomatic QFT in dS space was constructed from an analytic two-point function for massive and \textit{mcc} scalar field by Bros et al. \cite{brgamo,brmo}. Then it was generalized to the spinor field \cite{brgamota} and massless vector field $K_\beta$ \cite{gagarota}. Here, we generalize the previous work to the massless vector field $K_\beta^{a}$ in the Yang-Mills theory. The free vector field equation for $K_\beta^{a}$ is the same as for the free massless vector field $K_\beta$ in dS space. Differences appear at the interaction level or for the source of the fields. On the classical level for the vector field $K_\beta$, the source originates in the electrically charged spinor fields, whereas in the present case, the source of the free vector field $K_\beta^{a}$ stems from the colored spinor fields and as well the vector fields $K_\beta^{b}$, with $a\neq b$, {\it i.e.} $K_\beta^{a}$ is not a source for itself on the classical level, which means that there is no vertex with two same-colored vector fields $K_\beta^{a}$.

In the quantization procedure of the non-abelian gauge theory, $K_\beta^{a}$, two types of \textit{mmc} scalar fields appear. The first type stands for the scalar and longitudinal modes of the free vector field, which is equivalent to the massless vector field $K_\beta$. These scalar fields are necessary for a covariant quantization of the free massless vector filed, and they appear in the Gupta-Bleuler triplet construction, {\it i.e.} the free vector field operator transforms under a specific indecomposable representation of the dS group \cite{gagarota}. The other type appears as ghost fields, and they are needed to absorb the singularity of the Feynman path integral. These two types of auxiliary fields cannot propagate in the physical space or the external line of the Feynman diagrams. The first one is completely decoupled from the theory due to the conserved current, and it does not propagate in the internal line of the Feynman diagrams. However, the second one (ghost fields) interacts with the vector fields, which cannot be decoupled from the theory. Therefore it can propagate in the internal line of the Feynman diagrams, and it must be handled with precision. 

Let us first examine the free massless vector field. The standard approach is based on the following massless bi-tensor two-point function:
\begin{equation}
\label{fmvf}
\mathcal{ W}_{\beta \beta'}^{aa'}(x,x')=\langle
\Omega \mid K_{\beta}^{a}(x)K_{\beta'}^{a'}(x') \mid \Omega\rangle\, ,
\end{equation}
where $|\Omega\rangle $ can be chosen as the Bunch-Davies vacuum state and $x, x' \in X_H$.
These functions have to satisfy the following requirements:
\begin{enumerate}
\item[a)] {\bf Indefinite sesquilinear form:}
For any $5$-component test function $\big(f_\beta^a\big) \in \mathcal{ D}(X_H)$, we have an
indefinite sesquilinear form that is defined by
\begin{equation} \int _{X_H \times X_H}
f^{*\beta}_a(x)\mathcal{ W}_{\beta \beta'}^{aa'}(x,x')
f^{\beta'}_{a'}(x')\ud \sigma(x)\ud \sigma(x')\, ,\end{equation}
where $ f^*$ is the complex conjugate of $f$ \cite{brmo}. $\mathcal{ D}(X_H)$ is the
space of $C^\infty$ functions with compact support in $X_H$ and with values in $\setC^5$.
\item[b)] {\bf dS Covariance:}
The two-point function satisfies the covariance property
\begin{equation}R^{-1}\mathcal{ W} (R x,R x')R=\mathcal{
W}(x,x')\, .
\end{equation}
where $R \in \mathrm{SO}_0(1,4)$.
\item[c)] {\bf Locality:}
For every space-like separated pair $(x,x')$, {\it i.e.} $x\cdot
x'>-H^{-2}$ or $\mathcal{ Z}<1$, we have:
\begin{equation}\mathcal{ W}_{\beta \beta'}^{aa'}(x,x')=\mathcal{
W}_{ \beta' \beta}^{a'a}(x',x)\, .
\end{equation}
\item[d)] {\bf Normal analyticity:}
$\mathcal{ W}_{\beta \beta' }^{aa'}(x,x')$ is the boundary value (in the
sense of distributions) of an analytic two-point function $W_{\beta
\beta'}^{aa'}(z,z')$. The analyticity properties of the tensor two-point function in the tuboid $ \mathcal{ T}_{12}=\{
(z,z');\;\; z\in \mathcal{ T}^+,z'\in \mathcal{ T}^- \} $ follow from the
analyticity properties of the dS plane waves solutions \cite{brmo}.
\item[e)] {\bf Transversality:} The transversality with respect to $x$
and $x'$ is guaranteed since the dS modes solutions are transverse
by construction,
\begin{equation} x\cdot \mathcal{ W}(x,x')=0=x'\cdot \mathcal{
W}(x,x')\, .\end{equation}
\end{enumerate}
The free vector field two-point function reads as:
$$ \mathcal{ W}_{\beta \beta'}^{aa'}(x,x')=\delta^{aa'} D_{\beta
\beta'}(x,x',\partial_{x})\mathcal{ W}_{mcc}(x,x')\, ,$$
where $D_{\beta \beta'}(x,x',\partial_{x})$ is the bi-tensor differential operator (\ref{do}) and $\mathcal{ W}_{mcc}$ is the two-point function of the \textit{mcc} scalar field (\ref{stpci2}). These functions must entirely encode the theory of the generalized free fields on dS space-time $X_H$
and allow us to build the quantum field operators. Nevertheless, this approach does not work for the Yang-Mills theory due to the appearance of ghost fields as a necessary part of the vector fields $K^a_\alpha$. In this theory, for fixing the gauge (\ref{gfghost}), the ghost field is necessary and appears in the Feynman propagators' internal line. Within this framework, an axiomatic quantum field theory for the scalar field \textit{mmc} does not exist and then an axiomatic quantum Yang-Mills theory cannot be constructed.

Suppose we can ignore this difficulty and use the standard approach to quantize Yang-Mills theory. We then face serious difficulties with this construction, namely, a unique dS invariant vacuum state does not exist with this theory, the regularization of the IR divergence breaks the dS invariant, and it is not clear whether the theory is renormalizable or not at all orders of the perturbation due to the breaking of the dS invariance.

\subsection{Krein approach}
\label{kreinapp}
We focus on the quantization of the \textit{mmc} scalar fields since for the other fields the generalization is straightforward. In Krein space quantization, the decomposition of the field operator into positive and negative norm parts reads as \cite{gareta,gahure}:
\b \Phi_{\mathrm{mmc}}(x)\equiv\frac{1}{\sqrt{2}}\left[ \varphi_p(x)+\varphi_n(x)\right]\, ,\e
where
\b \varphi_p(x)=\sum_{k\geq 0} a_{k}\phi_{k}(x)+h.c.\, ,\;\;
\varphi_n(x)=\sum_{k \geq 0} b_{k}\phi^*_k(x)+h.c.\, .\e
The positive mode part, $\varphi_p(x)$, is the scalar field as was used by Allen.
The crucial departure from the standard QFT based on CCR lies in the
following commutation relations:
\b [a_{k},a^{\dag} _{k'}]= \delta_{kk'}\, ,\;\;
\;\;[b_{k},b^{\dag} _{k'}]= -\delta_{kk'}\, .\e
It is important to note that we now have a Krein-Fock vacuum state. Within this Krein space framework, the two-point function is the imaginary part of the two-point function of the positive mode solutions and it reads as \cite{ta01}:
\begin{equation}
 \label{tfsmin} 
 \begin{split}
\mathcal{ W}_{\mathrm{mmc}}^k(x,x')&=\ii \;\mathrm{Im} \left(\mathcal{ W}^{\mathrm{AF}}_{mmc}(x,x')\right)=\langle\Omega^k\mid \Phi_{\mathrm{mmc}}(x)\Phi_{\mathrm{mmc}}(x') \mid \Omega^k\rangle\\
 &=\frac{\ii H^2}{8\pi} \epsilon (x^0-x'^0)[\delta(1-\mathcal{ Z}(x,x'))-\theta (\mathcal{ Z}(x,x')-1)]\, , 
\end{split}
\end{equation}
where $\theta$ is the Heaviside step function, and $\mid \Omega^k>$ is the Krein-Fock vacuum state. This function is free of IR divergence, is dS invariant, and tends to a constant quantity in the large separated points limit. It is built on an indefinite metric space, which is not positive definite.

The ``Feynman'' propagator or  time-ordered product propagator in Krein space quantization reads as:
$$ G_T^k(x,x')=-\ii \langle\Omega^k\mid T\Phi_{\mathrm{mmc}}(x)\Phi_{\mathrm{mmc}}(x') \mid \Omega^k\rangle$$
\b \label{gfmin} =\frac{ H^2}{8\pi} \left[\delta(1-\mathcal{ Z}(x,x'))-\theta (\mathcal{ Z}(x,x')-1)\right]\, ,\e
where $T$ is the time ordering operator relative to the ambient coordinate $x^0$ \cite{brmo}. This function is causal and free from any IR and ultraviolet divergence. It owns light-cone singularity. This singularity can be absorbed in the fluctuation of the quantum metric and does not appear in the calculation of the vacuum expectation value of the physical quantity \cite{ta02}. Then, considering the quantum field theory in Krein space quantization with quantum metric fluctuations included, it is proved that all singular behaviors of the free scalar Green functions are removed. The new function $\mathcal{ Z}$ is introduced in order to take into account  the metric fluctuations: 
\b \label{linear} g_{\mu\nu}=g_{\mu\nu}^{dS}+h_{\mu\nu},\;\; |h|<|g^{dS}|\, .\e
It expands as  $\mathcal{ Z}= \mathcal{ Z}_0+\mathcal{ Z}_1+\cdots$, where $\mathcal{ Z}_0$ is given by the equation (\ref{gedi}) and $\mathcal{ Z}_1$ is the first-order perturbation due to the gravitational wave $h_{\mu\nu}$. In order to take into account the fluctuations of the quantum metric (see Appendix \ref{A}) and in analogy with the minkowskian case, the Green function (\ref{gfmin}) must be replaced with the following equation \cite{ta02}:
\b \label{gfminmf} \langle G_T^k(x,x')\rangle \approx \frac{ H^2}{8\pi} \left[\frac{1}{\sqrt{\pi (1-\langle\mathcal{ Z}_1\rangle )^2} }
\exp\left(-\frac{(1-\mathcal{ Z}_0)^2}{(1-\langle\mathcal{ Z}_1\rangle )^2} \right)-\theta (\mathcal{ Z}_0(x,x')-1)\right]\, .\e
 The expectation value $\langle\mathcal{ Z}_1\rangle \neq 1$ is related to the effect of first order quantum metric fluctuations. It determines the density of gravitons \cite{for3}. More precisely $\langle\mathcal{ Z}_1\rangle$ in \eqref{gfminmf} features the quantum metric fluctuations due to the quantum field operator $h_{\mu\nu}$. In the limit $\langle\mathcal{ Z}_1\rangle \longrightarrow 1$ the delta function is reobtained:
\b \lim_{a \longrightarrow 0} \frac{1}{\sqrt{\pi a^2}}
\exp\left(-\frac{x^2}{a^2} \right)=\delta(x)\, .\e
This means that the light-cone fluctuations disappear in this limit.

The Krein approach is based on the massless two-point function (\ref{tfsmin}) and it must satisfy the following requirements (to be compared with those for the function \eqref{fmvf}) \cite{gareta,gahure,engapewa}:
\begin{enumerate}
\item[A)] {\bf Indefinite sesquilinear form:}
For any test function $f(x) \in \mathcal{ D}(X_H)$, we have an
indefinite sesquilinear form that is defined by
\begin{equation} \int _{X_H \times X_H}
f^*(x)\mathcal{ W}_{\mathrm{mmc}}^k(x,x')
f(x')\ud \sigma(x)\ud \sigma(x')\, ,\end{equation}
where $ f^*$ is the complex conjugate of $f$. $\mathcal{ D}(X_H)$ is the
space of $C^\infty$ functions with compact support in $X_H$ and with values in $\setC $.
\item[B)] {\bf dS Covariance:}
The two-point function satisfies the covariance property
\begin{equation}\mathcal{ W}_{\mathrm{mmc}}^k(R x,R x')=
\mathcal{ W}_{\mathrm{mmc}}^k(x,x')\, .
\end{equation}
where $R \in\mathrm{ SO}_0(1,4)$.
\item[C)] {\bf Locality:}
For every space-like separated pair $(x,x')$, {\it i.e.} $x\cdot
x'>-H^{-2}$ or $\mathcal{ Z}<1$, we have:
\begin{equation}\mathcal{ W}_{\mathrm{mmc}}^k(x,x')=\mathcal{ W}_{\mathrm{mmc}}^k(x',x)\, .
\end{equation}
\end{enumerate}
These properties are not sufficient to generalize the Krein space quantization to the quantum theory for interacting fields such as Yang-Mills. For this purpose, we need the analyticity properties of the vacuum expectation value of the field operators \cite{brmo,stwi}:
\b \label{vevt} \langle\Omega^k\mid T \,\phi(x_1)\phi(x_2) \cdots \phi(x_n) \mid \Omega^k\rangle \, .\e

Since there is not yet rigorous mathematical proof for the analyticity properties of the vacuum expectation value of the field operators in Krein space quantization, we use the Krein space quantization, including quantum metric fluctuations, as a method of quantum field regularization (for details, see \cite{ta02}). Then we impose a supplementary condition on the two-point function in view of  generalization to the quantum theory for  interacting fields:
\begin{enumerate}
\item[D)] {\bf Light-cone fluctuations:} The delta function of the internal line in the vacuum expectation value of the time-order product of field operators (\ref{vevt}), is regularised in the following way:
\b \label{repexp}
\delta(1-\mathcal{ Z}_0(x,x'))\, \mapsto \, \frac{1}{\sqrt{\pi (1-\langle\mathcal{ Z}_1\rangle )^2} }
\exp\left(-\frac{(1-\mathcal{ Z}_0)^2}{(1-\langle\mathcal{ Z}_1\rangle )^2} \right) .\e
\end{enumerate}
This condition D) supersedes the previous analyticity property d) requested for \eqref{fmvf}.
$\langle\mathcal{ Z}_1\rangle$ is a complex function of the space-time coordinates $x^\alpha$ \cite{for3}. We nevertheless have $\langle\mathcal{ Z}_1\rangle \neq 1$ due to the quantum fluctuations, and then the replacement \eqref{repexp} results in a regular function on the light-cone.

The generalization of the approach \textit{\`a la} Krein to spinor and vector fields is straightforward and can be constructed by replacing the two-point function with its imaginary parts and the condition D). Some of the advantages of this approach can be mentioned, namely,
\begin{itemize}
  \item existence of a unique Krein-Fock vacuum state,
  \item preservation of dS invariance,
  \item renormalizability of the theory at all perturbation orders,
  \item explanation of mass gap and color confinement issues in a covariant way, as is discussed in the next section.
\end{itemize}

If we take Condition D) as a {\it Light-cone fluctuations principle} in Krein space quantization, we get an axiomatic quantum Yang-Mills theory within the framework of the dS ambient space formalism, and this approach can explain the mass gap and color confinement problems in a covariant way. This principle makes sense since it is supported by physical evidence. The gravitational field exists, $g_{\mu\nu}=g_{\mu\nu}^{dS}+h_{\mu\nu}\,$ and at the Planck scale the quantum linear gravity is essential, $ \Delta h= \sqrt{\langle h\cdot h\rangle-\langle h\rangle^2}\neq 0\,$. Fluctuations of the dS light-cone are induced by quantum linear gravity, and then the Dirac delta function supported by the classical light-cone must be replaced along Condition D) \cite{ta02,for3}. Since this principle only changes the internal line of the Feynman diagrams, it leads to the regularization of the theory, and then $\langle\mathcal{ Z}_1\rangle$ can be considered as a regularization parameter. Through the renormalization of the theory, observable physical results are achieved \cite{ta02}.


\section{Mass gap and confinement}
\label{gapcolor}

The IR divergence of the massless minimally coupled scalar field resembles features of a massless field theory in $2$-dimensional space-time \cite{nakan}. The regularization of the infrared divergence in $2$-dimensional massless field theory results in a mass gap \cite{albe} and a confinement \cite{dcas}. Motivated by these facts, let us examine if a similar mechanism can be relevant to the Yang-Mills theory in dS space-time. Here two approaches for this problem, which were presented in the previous section, are discussed.

From the classical Lagrangian (\ref{spvegh}), there is no explanation for the mass gap and the color confinement in the dS Yang-Mills theory. On the QFT level, the effective Lagrangian or ``quantum Lagrangian'' can be established through the loop correction to the classical Lagrangian:
\b \mathcal{ L}_q=\mathcal{ L}_c+ \hbar \mathcal{ L}_1+ \hbar^2 \mathcal{ L}_2+\cdots \,.\e
In the one-loop approximation the appearance of ghost field propagators occurs in the Feynman diagrams. By considering the gluon self-energy diagram (or gluon mass term) the following expression appears in the effective action, which contains two vertices (each vertice has one vector field and two ghost fields):
\b \label{masgap} \cdots+ \int \ud\sigma(x)\ud\sigma(x')T \left[{\Phi^a} ^\dag(x) C^{abc}\partial^\top \cdot K^c(x) \Phi^b(x) \right]\left[{\Phi^{a'}} ^\dag(x') C^{a'b'c'}\partial'^\top \cdot K^{c'}(x') \Phi^{b'}(x') \right]+\cdots\,.\e
Having in view the finding of a mass term for the vector fields, we consider two vector fields that are connected with the two propagators of the \textit{mmc} scalar field. Let us go through the two different approaches discussed in the previous section.
\begin{enumerate}
\item[{\bf A)}] {Standard approach}:
With the dS flat coordinates used by Allen, the Green functions are calculated by using continuous sums of the type $\int_\mu^\Lambda \ud k f(k)$ \cite{al,alfol}, where $\mu$ is an IR cut-off or regulator, and $\Lambda$ is an ultraviolet cut-off. This regulator excludes the range $k\in [0 ,\mu]$ from the mode sum, so the zero modes' contribution is dropped. The introduction of the cut-off $\mu$ and $\Lambda$ in the propagator of the \textit{mmc} scalar field entails the appearance in Equation (\ref{masgap}) of a mass term of the following type for the gluon fields (see Appendix \ref{B}):
$$ \int \ud\sigma(x)\ud\sigma(x') M_{cc'}^{\alpha\alpha'}(\mu,\Lambda;x,x',\partial,\partial')K_\alpha^c(x)K_{\alpha'}^{c'}(x')\, .$$
The explicit form of this integral is not essential for our purpose here. Although this mass term breaks dS invariance, it can explain the short-range force of the strong interaction and the mass gap problem of the quantum Yang-Mills theory in dS space.
\item[{\bf B)}] {Krein approach}:
In this approach the propagator of the \textit{mmc} scalar fields is a regular function (\ref{gfminmf}), and by replacing it in equation (\ref{masgap}) give:
$$ \int d\sigma(x) \ud\sigma(x') M_{cc'}^{\alpha\alpha'}(H,\langle\mathcal{ Z}_1\rangle;x,x',\partial,\partial')K_\alpha^c(x)K_{\alpha'}^{c'}(x')\, .$$
In this method, the dS invariance is preserved, and one can explain the short-range force of the strong interaction and the mass gap problem. In this expression $H$ and $\langle\mathcal{ Z}_1\rangle$ play the role of the cut-off parameters.
\end{enumerate}

The behaviour of the two-point function for the massless vector field in the inflationary epoch, ($H\to $ large value), allows us to understand better the mass gap and the confinement problems in this limit. Let us here follow the standard approach. Let us pick up two points: $X^\mu=(t,\chi,\theta,\varphi)$ and $X'^\mu=(0,0,0,0)$ in terms of the global coordinates system (\ref{gcs}). Then in the limit of large $H$, we have: $ \mathcal{Z}=-H^2 x\cdot x'=\cos\chi \cosh Ht \approx \frac{1}{2}\cos\chi e^{Ht}\,$.
In this limit the two-point function of the conformally massless scalar field reads as:
\b \mathcal{ W}_{\mathrm{mcc}}(x,x')= \frac{H^2}{8\pi^2}\frac{1}{\frac{1}{2}\cos\chi e^{Ht}}= \frac{H^2}{4\pi^2 \cos\chi}e^{-Ht}\, .\e
Then the massless vector two-point function (\ref{vtfsg}) can be written in terms of the \textit{mcc} scalar two-point function as follows ($ \mathcal{Z} \neq 1$) \cite{gagarota}:
$$\mathcal{ W}_{\alpha \alpha'}(x,x')=\left[-\theta'_{\alpha'}\cdot \theta_\alpha\,\z \frac{\ud}{\ud\mathcal{ Z}}
+H^2(\theta'_{\alpha'}\cdot x)(\theta_\alpha\cdot x')\left(
2\frac{\ud}{\ud\mathcal{ Z}}+\mathcal{ Z}\frac{\ud^2}{\ud\mathcal{ Z}^2}\right)\right]\mathcal{ W}_{\mathrm{mcc}}(x,x')\, $$
$$ = \frac{H^2}{8\pi^2}\left( \eta_{\alpha\alpha'}+H^2 x'_\alpha x'_{\alpha'}+H^2 x_\alpha x_{\alpha'}-H^2 x_{\alpha}x'_{\alpha'} \mathcal{ Z} \right) \frac{\mathcal{ Z}}{(1-\mathcal{ Z})^2} $$
\b -\frac{H^4}{4\pi^2} \left(x'_\alpha x_{\alpha'}- x'_\alpha x'_{\alpha'}\mathcal{ Z}- x_\alpha x_{\alpha'}\mathcal{ Z} + x_{\alpha}x'_{\alpha'}\mathcal{ Z}^2\right)\left( \frac{\mathcal{ 1}}{(1-\mathcal{ Z})^2} + \frac{\mathcal{ \mathcal{ Z}}}{(1-\mathcal{ Z})^3} \right) \, . \e
At large values of $H$ the latter behaves as:
\b \lim_{H \rightarrow\infty} \mathcal{ W}_{\alpha \alpha'} (x,x')\Longrightarrow -\frac{H^4}{4\pi^2} x_\alpha x'_{\alpha'} \left( \frac{3\mathcal{ \mathcal{ Z}^2}}{2(1-\mathcal{ Z})^2} + \frac{\mathcal{ \mathcal{ Z}^3}}{(1-\mathcal{ Z})^3} \right)\approx -\frac{H^4}{8\pi^2} x_\alpha x'_{\alpha'} \, .\e
Terms linear in the coordinates $x^{\alpha}$'s and ${x^{\prime}}^{\alpha}$'s respectively appear in this limit. Such a behavior of the two-point function may be used to explain the mass gap and the color confinement problems \cite{jawi}. In the null curvature limit ($H=0$) one can see that this behavior disappears as well.


\section{Conclusions}
\label{conclu}
The non-abelian gauge-invariant Lagrangian density is calculated in the dS ambient space formalism. We obtain that the ghost fields are the \textit{mmc} scalar fields. We have examined the possibility of an axiomatic QFT for the Yang-Mills theory within the framework of Krein space quantization.  Due to the regularization of IR divergences, a mass term is obtained for the gluon fields in the one-loop approximation. This mass term may be used to solve the mass gap and color confinement problems in dS Yang-Mills theory. We show that in the large curvature limit ($H\rightarrow \infty$), early universe, or inflationary time, a term linear in system coordinates appears in the vector two-point function. In a forthcoming paper, we plan to consider the cluster decomposition principle by using the dS covariance, locality, the mass gap, and the principle D) within our Krein approach.

\vskip 0.5 cm
\noindent {\bf{Acknowledgements}}: The authors are grateful to Edouard Brezin, Eric Huguet, Jean Iliopoulos, and Jacques Renaud for discussions. We are grateful to the referee for precise and valuable comments. One of the authors (MVT) would like to thank le Coll\`ege de France and l'Universit\'e de Paris for their financial support and hospitality.


\begin{appendix}

\setcounter{equation}{0}
\section{}\label{A}

The effect of linear quantum gravity on the scalar Green's function can be calculated by considering the interaction between two fields. For the sake of simplicity, the massless scalar field with minimal coupling to the gravitational field is considered and Minkowski space time is chosen as the background. In this case, the total classical action reads:
\b \label{grascal} S_c[g,\phi]= \int \sqrt{-g}\; \ud^4x \left(R+\frac{1}{2} g^{\mu\nu}\partial_\mu \phi \partial_\nu \phi \right)\,,\e
where $R$ is the scalar curvature. In the linear approximation of the gravitational field ($g_{\mu\nu}=\eta_{\mu\nu}+h_{\mu\nu},\; |h|<|\eta|$) and with the definition $\sqrt{-g}\equiv 1+f(h)$, where $f$ is a differentiable function, we have \cite{car}:
\begin{equation}
\label{linearscal}
\begin{split}
S_c[g,\phi]&= \int \left(1+f(h)\right) \ud^4 x \left(\frac{1}{2}\left[ (\partial_\mu h^{\mu\nu})(\partial_\nu h)
-(\partial_\mu h^{\rho\sigma})(\partial_\rho h^\mu_\sigma)\right.\right.
\\
& \left.\left.+\frac{1}{2}\eta^{\mu\nu} (\partial_\mu h^{\rho\sigma})(\partial_\nu h_{\rho\sigma})
-\frac{1}{2}\eta^{\mu\nu}(\partial_\mu h)(\partial_\nu h)
\right] +\frac{1}{2} \left(\eta^{\mu\nu}+h^{\mu\nu}\right)\partial_\mu \phi \partial_\nu \phi \right)\, .
\end{split}
\end{equation}
Consider the term $f(h)\eta^{\mu\nu}\partial_\mu \phi \partial_\nu \phi $ appearing  in the Lagrangian density. On the quantum metric fluctuation level it may be written as:
\b \label{conterterm}  f(\langle h^2\rangle)\eta^{\mu\nu}\partial_\mu \phi \partial_\nu \phi\, .\e
If we assume that $\langle h^2\rangle$ is constant, this term may be interpreted as a counterterm for the scalar field. It modifies the scalar Green function and the factor $f\equiv Z_\phi$ plays the role of a wave function renormalization parameter.
The other quantum effects show up through the quantum effective action, which can be written as:
$$ S_q[g,\phi]= S_c[g,\phi]+\hbar S^{(1)}[g,\phi]+ \cdots \,,$$
where $S^{(1)}$ is the one-loop approximation.

\section{} \label{B}

In this appendix a brief procedure for calculating $M_{cc'}^{\alpha\alpha'}(\mu,\Lambda;x,x',\partial,\partial')$ is presented. We start out with the following expression,
$$ T \left[{\Phi^a} ^\dag(x) C^{abc}\partial^\top \cdot K^c(x) \Phi^b(x) \right]\left[{\Phi^{a'}} ^\dag(x') C^{a'b'c'}\partial'^\top \cdot K^{c'}(x') \Phi^{b'}(x') \right].$$ 
It can be divided into four terms where, for simplicity, solely one of them is considered:
$$ C^{abc}C^{a'b'c'}T \left[{\Phi^a}^\dag (x)\Phi^b(x) \partial^{\top\alpha}  K^c_\alpha(x)  \right]\left[{\Phi^{a'}}^\dag (x')\Phi^{b'}(x') \partial'^{\top\alpha'}  K^{c'}_{\alpha'}(x')  \right].$$
Since we are looking for a mass term for the vector fields, then the vector fields do not participate in the time order product. For scalar fields, the time-ordered product and vacuum expectation value result in the following expression:
$$ C^{abc}C^{a'b'c'}\delta^a_{b'} \delta^b_{a'} \left[ G_T(x,x')\right]^2\partial^{\top\alpha} \partial'^{\top\alpha'} K^c_\alpha(x) K^{c'}_{\alpha'}(x')\, .$$
It is part of the expression $M_{cc'}^{\alpha\alpha'}$ in which $G_T(x,x')$ is the time-ordered scalar Green function. The latter is different from what is yielded by the standard model and by the Krein approach.

\end{appendix}

\end{document}